# Beyond Trade Openness: Network-Based Evidence on African Economic Integration

Tekilu Tadesse Choramo[1,2], Jemal Abafita[2], Yerali Gandica[4,5,6] and Luis E C Rocha[1,3]
[1]Department of Economics, Ghent University, Ghent, Belgium
[2] Department of Economics, Jimma University, Jimma, Ethiopia
[3]Department of Physics and Astronomy, Ghent University, Ghent, Belgium
[4]Valencian International University (VIU), Valencia, Spain
[5]Universidad de Nebrija, Madrid, Spain
[6]Department of Economic Analysis: Quantitative Economics, Universidad Autónoma de Madrid, Madrid, Spain

**Abstract**

*This paper develops a network-based methodology for measuring economic integration in Africa. Conventional indicators such as trade openness and intra-regional trade shares capture the volume of trade but not countries' structural roles within the continental trade system. Using bilateral intra-African trade data for 2000–2019, we construct annual directed trade networks and measure countries' direct connectivity, recursive influence, brokerage role, and core embeddedness using weighted degree, PageRank, betweenness and random-walk betweenness, clustering, and k-core network indicators. We further benchmark observed network positions against a maximum-entropy null model to identify countries whose centrality exceeds that expected from their trade intensity alone. The results show a gradual densification of intra-African trade and the persistence of a core–periphery structure led by a small group of highly connected economies. Dynamic panel estimates indicate that economic development, infrastructure, institutional quality, human capital, regional trade agreements, and FDI are associated with stronger structural integration, whereas trade costs and overlapping regional memberships weaken several dimensions of network position. The findings suggest that African integration depends not only on expanding trade volumes but also on improving countries' positions within continental trade.*

## 1. Introduction

Africa remains on the periphery of the global economy, with a relatively small and declining share of world trade. According to the International Monetary Fund's (IMF) direction of trade (DOT) statistics, only about 10–15% of African trade is conducted between African countries. Over 80% of African exports go to countries outside of Africa. Similarly, despite Africa's potential to meet a larger share of its import demand internally, over 90% of imports come from outside the continent, indicating that intra-regional trade remains below its potential(Admassu, 2019). Economists concur that expanding intra-continental trade is one of Africa's most effective strategies to support economic growth and development (Olney, 2020; Admasu, 2019).

Despite persistent poverty, African countries have experienced growth, with an average growth rate of 4.7% between 2000 and 2019, largely due to strengthened intra-regional trade (UNECA, 2019). Debates continue on how to achieve sustainable economic growth in Africa, given many countries' reliance on foreign aid rather than trade (Moy, 2009). Many African countries have become increasingly open to leveraging preferential trade arrangements within their regional economic communities (Admassu, 2019). The African Continental Free Trade Area Agreement (AfCFTA), signed in 2018, aims to deepen economic integration by increasing intra-African trade, strengthening production–export linkages, creating jobs, and reducing members' exposure to commodity price volatility. As an economic bloc, the African Union (AU) comprises fifty-four member states and recognizes eight regional economic communities (hereafter RECs), some with overlapping memberships (see Table A.11 in the appendix). Yet, the continent remains weakly integrated into global trade networks.

Empirical studies of Africa have used different measures of integration, including intra-regional trade, regional trade agreements, trade openness, and gravity measures (Rekiso, 2017; Vhumbun, 2019; Admassu, 2019; Gammadigbe, 2021). These studies often rely on aggregate or standard macroeconomic indicators, which may not capture the complexity of interactions among trading partners. As a consequence, there has been increasing interest in using more advanced techniques from network science to measure economic integration by quantifying the interconnectedness of countries and the complex relationships between them (Garlaschelli et al., 2007; Fagiolo et al., 2008; Iapadre and Tajoli, 2014; Herman, 2022). Complex network theory provides a framework for analyzing interactions among countries when trade relationships form dense and interdependent systems (Kali and Reyes, 2007; Kali et al., 2007). Within this framework, countries are represented as nodes, while the edges between them correspond to bilateral trade relationships, typically measured by trade volume. The position of each country within the trade network is evaluated using various centrality measures, which capture different dimensions of a country's importance or influence in the system.

Figure 1 illustrate show network-based measures capture dimensions of integration that traditional trade indicators miss. Panel (a) represents trade openness: node 1 has many direct trade links, indicating strong bilateral trade activity. However, openness does not indicate whether these partners are themselves influential, whether the country connects otherwise-separate parts of the network, or whether it belongs to the dense core of the system. Panel (b) illustrates PageRank centrality, where node 1 is important because it is connected to other well-connected nodes. In trade terms, PageRank captures links to influential trading partners. Panel (c) illustrates betweenness centrality, with node 4 acting as a broker between weakly connected parts of the network. Panel (d) illustrates k-core centrality, where node 1 belongs to the innermost core and is embedded in a dense group of mutually connected countries. Network measures, therefore, move beyond bilateral trade intensity by distinguishing direct connectivity, influence, brokerage, and core embeddedness within the continental trade network. In this sense, integration is not only a question of how much countries trade, but also of where they sit in the structure of continental trade.

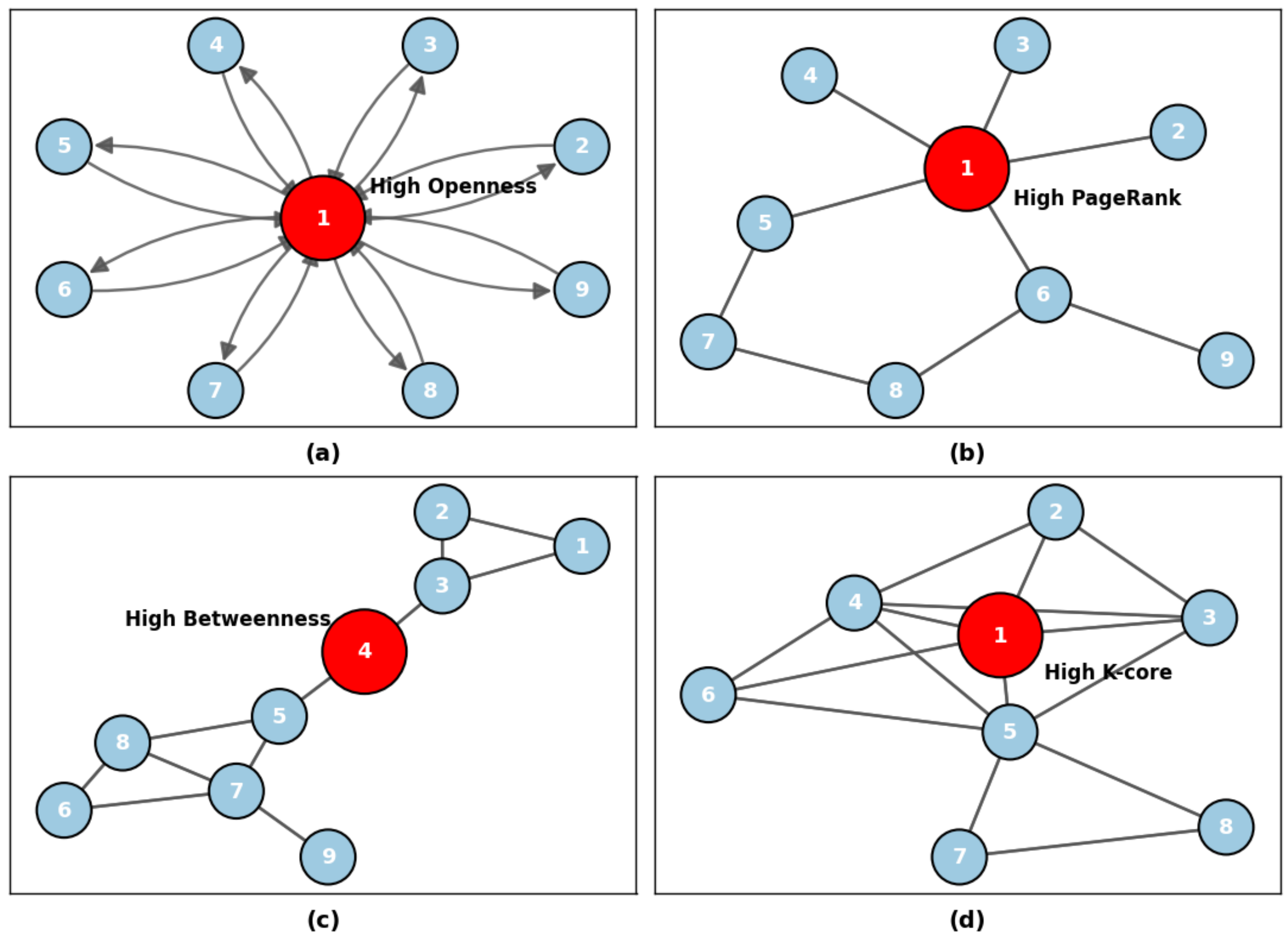

Figure 1: Model networks illustrating the most central nodes estimated from different centrality metrics. (a) openness (related to trade partners), (b) PageRank (related to prestige), (c) betweenness (brokerage), (d) k-core (about embeddedness). Node size indicates centrality values. The figure illustrates the conceptual meaning rather than empirical magnitudes.

However, few studies explicitly link trade-network formation to macroeconomic determinants (De Benedictis and Tajoli, 2011; De Lombaerde et al., 2018; Chong et al., 2019; Yuan et al., 2022), while others used network centrality measures as determining factors for the growth and finance of a given country (Kali and Reyes (2007); Reyes et al. (2010); Duenas and Fagiolo (2013); De Benedictis and Tajoli (2018); Herman (2022)) without considering the standard econometric model. Furthermore, some studies focused on network position and innovation potential (Nepelski & De Prato, 2018), as well as on firm productivity and intermediate trade network centrality (Ayadi et al., 2024). However, the link between macroeconomic indicators and network centrality metrics as measures of economic integration remains underexplored. This paper proposes advanced network indicators to measure economic integration in Africa, in particular, PageRank, random-walk centrality, and k-core decomposition, to capture different dimensions of interconnectedness rather than growth and trade effects of network position (Newman, 2005).

This study combines African trade-network analysis with dynamic panel estimation using longitudinal data. By linking network structure with macroeconomic fundamentals, we propose that centrality metrics can serve as quantitative indicators of economic integration in Africa. We further explore the determinants of these positions through panel regression analysis using relevant macroeconomic variables. By applying complex network methods, this study captures the structural architecture of African regional integration, revealing the disproportionate influence of centrally connected countries and the structural constraints faced by peripheral members. Using a dynamic regression model, we estimate how macroeconomic variables shape countries' structural roles within the African trade network.

This paper makes three contributions. It measures African economic integration as a structural position within a continental trade network, rather than only as a trade-volume ratio. This allows us to distinguish direct trade intensity, influence through well-connected partners, brokerage between weakly connected countries, and embeddedness in the trade core. It also uses a maximum-entropy benchmark to separate genuine structural prominence from centrality mechanically driven by trade volume. Finally, it links network-based integration measures to macroeconomic and policy determinants using a dynamic panel framework. The analysis, therefore, moves beyond describing the structure of African trade and asks why some countries become central while others remain peripheral.

## 2. Stylized facts for intra-African trade

Africa accounts for 14.4% of the world's population and 21.2% of its territory, but only 2.7% of global trade over the past two decades (Fig. 2a). Although intra-African trade (compared to its share in the global trade) expanded from around 10.4% in 2000 to roughly 17.8% in 2017 (Fig. 2a), it remains low compared with Europe (69%), Asia (59%), North America (47%), and South America (23%) (UNCTAD, 2021). In March 2018, African countries signed a landmark African Continental Free Trade Area Agreement (AfCFTA) to boost intra-African trade by removing tariffs and liberalizing trade. Intra-African trade remains weakly diversified, partly because many countries export similar primary commodities and have limited capacity to produce higher value-added goods (Olney, 2020).

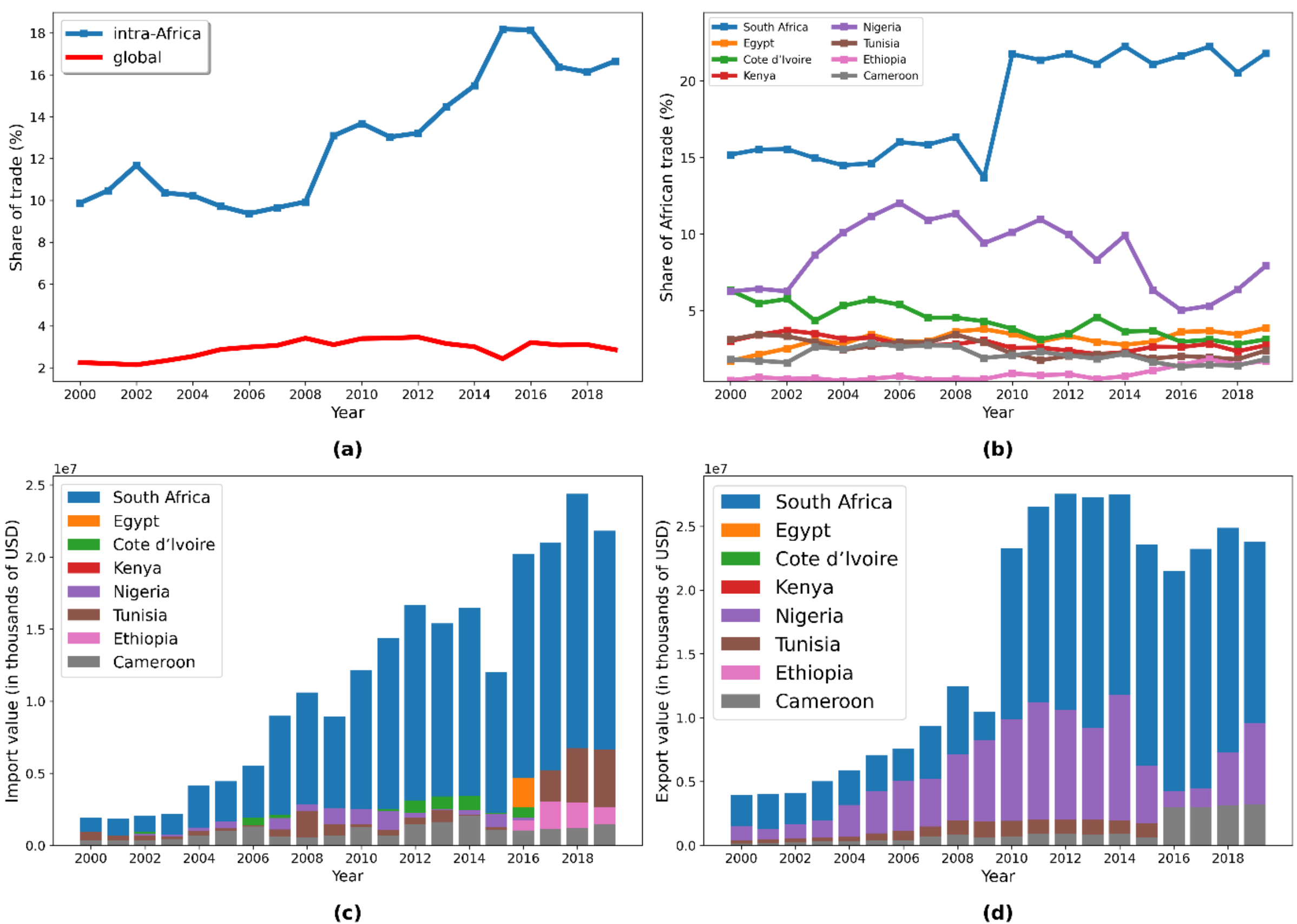

Figure 2: African trade structure and trade volume. (a) Intra-African and global trade share in the period 2000-2019, (b) intra-African trade share for eight selected countries, (c) total import value (in thousands of USD) for eight selected countries, and (d) total export value (in thousands of USD) for eight selected African countries.

Figure 2a shows that intra-African trade increased over the past 20 years. Although South Africa remained the largest participant after 2008, the rise in intra-African trade appears to reflect broader regional integration rather than the performance of a few dominant economies. Evidence from the United Nations Conference on Trade and Development (UNCTAD, 2012) suggests that trade participation expanded across a wider range of African countries through the strengthening of regional economic communities and the growth of regional market integration. This is reflected in the increasing importance of regional markets, greater diversification of intra-African exports, and the emergence of new trade linkages among African economies. These trends provide the context for the AfCFTA and its goal of reducing barriers to intra-African trade.

Figure 2b shows that South Africa consistently accounted for the largest share of African trade throughout the period, rising from approximately 15% in the early 2000s to over 22% after 2010, then declining slightly to around 20% by 2019. Nigeria was the second-largest contributor, with its share rising from about 6% in 2000 to a peak of nearly 12% in 2006–2007, then gradually declining to around 5% by the end of the sample period. Figures 2(c) and (d) present the evolution of imports and exports for selected African countries across geographic regions, which together account for more than 53% of total intra-African trade between 2000 and 2019. South Africa records the highest trade values throughout the period, maintaining a substantial lead over the other countries in both imports and exports. Import values generally trend upward, particularly after the mid-2000s, with notable increases in South Africa, Egypt, Nigeria, and Tunisia. Export performance also improved over the sample period, although growth patterns varied across countries. Nigeria emerged as the second-largest exporter, showing considerable expansion in export values during the period. Moreover, the declining relative share of some dominant countries suggests a gradual redistribution of trade activity across a broader group of African economies. The figures indicate a substantial increase in trade activity among the selected African economies over the study period. However, trade volumes remain highly concentrated in a few large economies, particularly South Africa and Nigeria.

## 3. Literature Review and Hypothesis Construction

Early contributions used network analysis to characterize the structure of international trade systems. Barigozzi (2005), Garlaschelli and Loffredo (2005), and Barigozzi et al. (2010) documented the non-random topology and evolving structure of trade networks, while Kali and Reyes (2007) demonstrated that countries' positions within the network are economically meaningful and related to the dynamics of globalization. Building on these foundations, De Benedictis and Tajoli (2011, 2018) and Fagiolo (2016) provided systematic frameworks for analyzing global and regional trade networks, highlighting the relevance of centrality, clustering, and core–periphery structures for understanding integration beyond bilateral trade flows. Subsequent studies documented common stylized facts across regions and periods and used network-based metrics to explain economic outcomes or regional integration patterns (Kastelle et al., 2006; Fagiolo, 2007; Fagiolo et al., 2008; Reyes et al., 2009; Piccardi and Tajoli, 2012; Iapadre and Tajoli, 2014; Nguyen et al., 2016; Beaton et al., 2017; Gandica et al., 2020; Yuan et al., 2021; Ayadi et al., 2024). However, this literature largely describes network structures or links network position to outcomes, leaving open the question of which macroeconomic and policy factors drive countries' structural integration within continental trade networks, particularly in Africa.

Previous research showed that high-performing Asian economies have moved from the periphery of the trade network toward its core, whereas Latin American integration has remained stagnant, according to studies by Reyes et al. (2007, 2010). Iapadre and Tajoli (2014) used network centrality measures to analyze emerging countries and trade regionalization and found that global integration forces dominate regional integration dynamics. According to Borgatti (2005), betweenness centrality has a strategic implication for regional integration because a country can exploit its position based on its comparative advantage. Andal (2017), on the other hand, used closeness and eigenvector centrality measures to identify the position of the Asia-Pacific region as an indicator of economic integration. Similarly, Nguyen et al. (2016) used network centrality to analyze the characteristics of ASEAN+3's trade and FDI integration. Beaton et al. (2017) examined regional and global integration in Latin America and the Caribbean and showed that countries are well connected in terms of trade links, but their trade intensity remains low. Reyes et al. (2009) proposed that random-walk betweenness centrality (RWBC) could better explain the varying levels of economic integration among countries with the same level of trade openness, indicating the inadequacy of traditional measures of economic integration.

Regarding the African continent, the United Nations Conference on Trade and Development (UNCTAD) used network analysis in its 2018 report to examine examines how Africa can transition from fragmented RECs to a unified AfCFTA to boost intra-African trade and structural transformation. However, this analysis only used data from a single year and centrality measures, which are unlikely to provide a comprehensive understanding of the intra-African trade network. Zhang and Batinge (2021) used social network analysis to investigate the patterns of intra-African trade between 2002, 2012, and 2017. Their study confirmed the previously proposed core-periphery structure and small-world phenomenon of African trade networks and further identified that the network has become denser over time.

**The influence of Macroeconomic variables on network-based integration measures**

A country's degree of economic development can promote economic integration in various ways (Akbari, 2021). Different studies have tried to identify the major determinant of network formation and its correlation with network structure (De Benedictis and Tajoli, 2011; Yuan et al., 2021). A more central position in the network is occupied by the countries with larger and wealthier economies, and vice versa (Garlaschelli & Loffredo, 2005; Kali & Reyes, 2007; Fagiolo et al., 2009; Jiang & Tamang, 2020). These countries have been found to remain stable in terms of ranked centrality, with exponential growth over time (Fagiolo et al., 2009). Criscuolo and Timmis (2018) and Ayadi et al. (2024) found evidence that the Global Value Chain (GVC) centrality is positively associated with productivity, suggesting that firms that are more connected to the GVC are more productive. More recently, Adarov (2021) discovered a positive correlation between PageRank centrality in the GVC and FDI networks and a country's economic size, as determined by real GDP. In contrast, GDP growth affects degree centrality in the FDI network (Jiang & Tamang, 2020) and in the agricultural trade network (Sun et al., 2022). Zhang et al. (2022) and Arif et al. (2021) confirmed that real GDP is a prominent determinant of a country's central position in the case of an international energy trade network and FDI network, respectively.

The literature suggests that regional trade agreements (RTAs) impact different dimensions of centrality measures in a network. For example, Jiménez-García and Rodríguez (2022) quantified the impact of 103 bilateral RTAs on bilateral trade flows and found that RTAs have a significant positive effect on the betweenness centrality measure of the countries involved. Similarly, Sada et al. (2022) analyzed the political distance network among Asian countries and found that RTAs have a significant positive impact on closeness and betweenness centrality measures. Basile et al. (2018) also suggested that trade costs shape intra-regional trade network structures.

Based on the comparative advantage argument, countries with more skilled labor are more productive and better able to trade with other countries. Most studies suggest that human capital is statistically significant in determining the indirect effect of networks (Bari & Jayanthakumaran, 2021) and that developing economies trade more with trading partners when those partners have a similarly skilled labor force at lower cost (Regolo, 2013).

Adarov (2021) provides evidence that labor productivity, a measure of human capital endowment, is positively associated with higher GVC PageRank centrality, indicating that skilled labor is an essential component of competitiveness and can place a country in a central position in trading. From a global investment network perspective, Bolívar et al. (2019) found that human capital endowment is a prime driver of FDI network centrality measures, particularly regarding lower labor costs.

Zhang and Batinge (2021) found that institutional factors and trade network formation are positively correlated in a trading network. By maintaining transparency, institutions' efficacy has been shown to reduce trade costs, which may motivate exporting based on comparative advantage (Abban, 2020b; Levchenko, 2013). Fracasso et al. (2018) confirmed that institutions' quality, measured by the democracy index, is positively associated with out-degree in the oil trade network. Recent studies have shown that institutional quality, as measured by political stability, is positively associated with GVC network connectivity in terms of PageRank centrality from FDI and value chain networks (Adarov, 2021) and degree centrality in the FDI network (Blonigen and Piger, 2014; Jiang and Tamang, 2020). Domestic institutions significantly impact value chain integration through RTA implementation and national specialization channels (Kügler et al., 2020).

The clustering coefficient measures the local connectivity, whereas the k-core considers the network's overall structure. This assumes that each network has a core, and nodes close to the core are the most influential (Qiu et al., 2021). Meliciani et al. (2022) suggest that European regions with higher rates of innovation and faster economic growth are more in the network core than those surrounded by other regions with strong connections (i.e., higher clustering coefficients). Consistent with the global trade network, the African trade network is expected to have a core-periphery structure (De Benedictis & Tajoli, 2011; Fagiolo et al., 2008). Chen et al. (2016) suggest that a country's competitiveness is associated with an increase in the k-core in the global natural gas trade network. The empirical evidence also found a positive correlation between the clustering coefficient and real per capita GDP (Antonietti et al., 2021) and regional agreements, which tend to encourage larger trade flows among trade network members (Zhang et al., 2014).

Accordingly, the study formulates three main hypotheses for empirical testing:

H1: African trade integration exhibits a persistent but evolving core–periphery structure.

H2: Development, institutions, infrastructure, human capital, RTAs, and FDI increase countries' structural integration within the African trade network.

H3: Trade costs and overlapping REC memberships weaken countries' structural integration within the African trade network.

## 3. Methods

### 3.1. African economy as a complex trade network

International trade between African countries is represented as a network. A network $H = (N, E)$ is defined by a set of N nodes and a set of E edges. In the trade network, each country is represented by a node $i$, and trade is represented by a directed edge $(i, j)$ from the exporter country $i$ to the importer country $j$. The adjacency matrix $A(t) = \{a_{ij}\}$ indicates whether there is trade$\left(a_{ij} = 1\right)$ or not $\left(a_{ij} = 0\right)$ between countries $i$ and j in year t. The trade flow is represented by the weighted adjacency matrix, $W(t)$, where $w_{ij}$ represents the total value of trade (in USD) from i to j in year t.

### 3.2. Network-based indicators of economic integration

We group the network-based measures into three categories that capture distinct dimensions of economic integration. Table A.7 in the appendix summarizes each indicator and the integration dimension it captures[1]. The first category of integration indicators comprises network measures based on a given country's direct trade partnerships. The degree centrality measures the number of trade partners normalized by the number of potential partners. The in-degree $k_i^{in}$ (eq. 1) and out-degree $k_i^{out}$ (eq. 2) represent, respectively, the number of incoming (import partner) and the number of outgoing (export partner) edges of the country $i$.

$$k_i^{in} = \frac{1}{N-1}\sum_{i=1}^{N} a_{ji} \qquad (1)$$

$$k_i^{out} = \frac{1}{N-1}\sum_{i=1}^{N} a_{ij} \qquad (2)$$

The weighted in-degree (eq. 3) and out-degree (eq. 4) complement the measures above by adding the value of trade (in USD) on the edges.

$$S_i^{in} = \frac{1}{N-1}\sum_{i=1}^{N} w_{ji} \qquad (3)$$

[1] There are also details on some of the basic network metrics, such as density, average clustering coefficient, and average degree of the network. The measures are normalized following standard practice, which ensures comparability across countries within the same network and across different years. Consequently, the centrality indicators are meaningful in terms of relative position and are directly comparable across African economies in our sample.

$$S_i^{out} = \frac{1}{N-1} \sum_{i=1}^{N} w_{ij} \qquad (4)$$

PageRank measures recursive importance in a network (eq. 5). PageRank is the eigenvector associated with the largest eigenvalue of the transition matrix (where d is conventionally set to 0.85). For a weighted directed network, $a_{ij}$ may be replaced by the weighted edges $w_{ij}$ and thus the out-degree $k_j^{out}$ by the out-weighted-degree $S_j^{out}$.

$$PR_i = \frac{1-d}{N} + d \sum_j \frac{a_{ij}}{k_j^{out}} PR_j \qquad (5)$$

The second category of indicators involves clustering. The clustering coefficient (cc) measures the local connectivity of a node and its trade partners (eq. 6) (Onnela et al., 2005; Fagiolo et al., 2010). A higher clustering coefficient indicates that a country's trade partners are also trading with each other. The clustering coefficient may incorporate edge weights to account for the level of trade between partners. The increase in the clustering coefficient has more implications for a country's integration within the group than between the groups in the network (Yuan et al., 2021; Reyes et al., 2010; Kali & Reyes, 2007).

$$cc_i = \frac{2e_i}{k_i(k_i-1)} \qquad (6)$$

where $e_i$ is the number of trade relationships between the trade partners of a country i.

Closeness centrality measures the distance between any two countries i and j in the African trade network. Closeness centrality reflects how easily a country can connect to the rest of the trade network through its trade relationships. Countries with high closeness centrality reach other trading partners through shorter trade paths, indicating faster access to markets and information flows. It reflects the efficiency of a country's integration within the trade network, rather than geographic distance. It is specified as:

$$C(i) = \left[ \frac{1}{N-1} \sum_{j \neq i} d(j,i) \right]^{-1} \qquad (7)$$

where $d(j,i)$ is the average length of the shortest paths between i and all the other nodes.

The k-core (or k-shell) decomposition identifies core and peripheral countries based on the density of their trade linkages. K-core shows whether a country is part of the core group of highly interconnected traders or on the periphery. The k-shell is obtained by iteratively removing nodes with degrees less than or equal to k. It first removes nodes with degree k=1 until no nodes remain with degree k=1. Then, the k-core of the removed nodes is set to k-core=1. In the following steps, nodes are iteratively removed with residual degree no greater than n to obtain the subsequent shells (Fig. 3). Nodes removed in step n have k-core = n (Seidman, 1983; Borgatti et al., 2018).

Low k-core values indicate peripheral positions, whereas high values indicate membership in the densely connected network core (Filho et al., 2018).

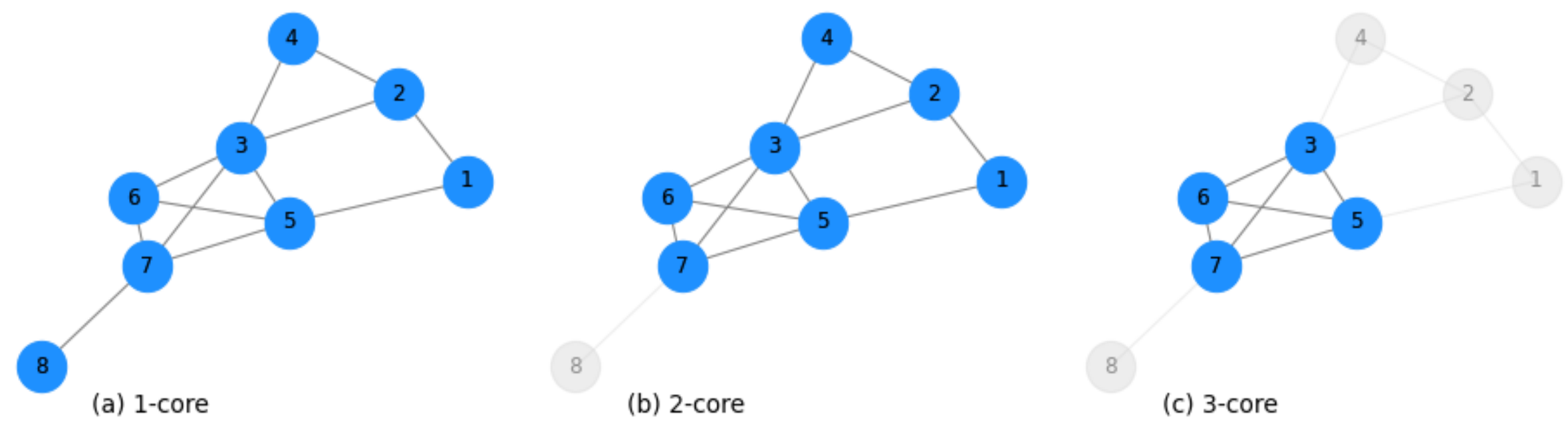

Figure 3. The corresponding (a) 1-core, (b) 2-core, and (c) 3-core (k-core =3) of a sample network with 8 nodes and 12 edges.

The third category of measures captures the brokerage potential of nodes. These indicators identify countries that lie on many trade routes and connect otherwise weakly linked partners. Betweenness-type measures capture this bridging role by quantifying how often a country sits on the shortest or most likely (random-walk) paths between other countries. It measures the sum, over all pairs of nodes in the network, of the fraction of shortest paths passing through a node $i$ with respect to all existing shortest paths between the same pair of nodes (eq. 8) (Goh et al., 2003). The shortest path $\sigma_{kj}$ between two nodes $k$ and $j$ is defined as the minimum number of edges necessary to connect these two nodes.

$$B_i = \sum_{(j,k)\ j \neq i \neq k} \frac{\sigma_{jik}}{\sigma_{jk}} \qquad (8)$$

Betweenness centrality captures network connectivity rather than potential flow over edges. Random-walk betweenness centrality (RWB) extends shortest-path betweenness by measuring how frequently a node is expected to be traversed when flows follow random walks between source-target pairs, which is relevant when trade interactions diffuse through multiple routes (Newman, 2005). Specifically, the number of times a random walk passes through a given node i along the path, averaged over all g and m, starting at node $j$ and ending at node $m$ is how the RWB of a given node $i$ as defined in (eq. 9).

$$RWB_i = \frac{1}{N(N-1)/2} \sum_{j \neq m} I_{jm}^{i} \qquad (9)$$

where $I_{jm}^{i}$ represents the number of times the node $i$ is passed during the random walk from the source node j to the target node m.

#### 3.2.1. Network Centrality Normalization Using a Maximum Entropy Null Model

To evaluate countries' structural integration within the intra-African trade network, we normalize all centrality indicators using a Maximum Entropy Null Model (MENM) (De Clerck et al, 2024). This approach constructs randomized benchmark networks that preserve each country's observed degree distribution in the binary case (i.e., number of trade links) and strength distribution in the weighted case (i.e., total trade volume), while randomizing all other topological properties. By comparing the observed centrality values with those generated from the null ensemble, we compute Z-scores that capture how statistically exceptional each country's position is relative to what would be expected by chance. Positive Z-scores indicate that a country's centrality is significantly higher than the random benchmark, while negative Z-scores indicate a lower-than-expected position relative to the benchmark. This normalization allows for meaningful cross-country and intertemporal comparisons, distinguishing genuine structural prominence from size-related effects. The detailed mathematical formulation and computational procedure of the maximum entropy null model are provided in Appendix A. In the empirical analysis, raw centrality indicators capture countries' observed network positions, while MENM-normalized Z-scores assess whether these positions are structurally exceptional relative to a randomized benchmark and are therefore, interpreted as evidence of structurally significant integration rather than as econometric variables.

### 3.3. Econometric Model

This study uses a dynamic panel model to investigate the macroeconomic drivers of network-based economic integration (Akbari et al., 2021; Fracasso et al., 2018; Bhattacharya et al., 2018). We explicitly address potential endogeneity in the explanatory variables. The dynamic version of the model is specified as:

$$Centrality_{it} = \alpha + \varphi Centrality_{it-1} + \beta X_{it} + \phi \pi_{it} + \gamma T_t + \varepsilon_i \, \mu_{it} \qquad (10)$$

where $Centrality_{it}$ denotes a network-based integration indicator, $X_{it}$ is a vector of macroeconomic controls, and $Centrality_{it-1}$ captures persistence in network positions. We include country and time effects, and $\pi_{it}$ captures multiple memberships in regional blocs and the 2007–2009 global financial crisis. The full model is specified as follows:

$$\begin{aligned} Centrality_{it} = \alpha_i &+ \beta_1 Centrality_{it-1} + \beta_2 GDP_{cit} + \beta_3 IQI_{it} + \beta_4 TC_{it} + \beta_5 pop_{it} + \beta_6 Infra_{it} \\ &+ \beta_7 HC_{it} + \beta_8 RTA_{it} + \beta_9 OFR_{it} + \beta_{10} FDI_{it} + \phi FinancialCrisis + \gamma T_t \\ &+ \mu_{it} \qquad (11) \end{aligned}$$

The dependent variable $Centrality_{it}$[2] is measured using the network indicators defined in Section 3.2. In contrast, the explanatory variables are real gross domestic product per capita (RGDPc), Human capital (HC), trade cost (TC), institutional quality index (IQI), Infrastructure index (Infra), population size (POP), Regional Trade Agreements (RTA), Foreign direct investment (FDI), overlapped membership in any of eight regional economic cooperation measured by overlap frequency ratio (OFR) and global financial crisis from 2007 to 2009 (Financial crisis). Except for k-core, variables were log-transformed, so the coefficients can be interpreted as percentage changes in a country's network position and macroeconomic characteristics.

---

[2] The trade network data is bilateral and varies over time, with observations recorded for each origin-destination country pair per year. The calculated centrality and non-centrality measures reflect the level of connectivity or integration of a given country within the network for a specific year. These measures have country and year dimensions, meaning they can be analyzed as a panel data structure.

The gross domestic product per capita (RGDPc) is included as a standard macroeconomic control variable to capture general differences in countries' economic conditions. The population (POP) aims to capture the potential of market size to boost regional trade for each country (Sabbagh et al., 2013). We expect the institutional quality variable to positively correlate with a country's network position. Six institutional quality indicators were assembled by the World Government Indicators (WGI) and published by the World Bank in 2020. We construct an institutional quality index using principal component analysis (PCA) based on six WGI measures. The index ranges from -2.5 to +2.5, where higher values indicate stronger governance. The infrastructure index (Infra) is constructed using principal component analysis (PCA) based on four indicators: fixed telephone lines per 100 people, air transport freight (ton-kilometers), energy use (kg of oil equivalent per capita), and electric power consumption (kWh per capita).

Unlike previous studies using policy outcome variables, we include regional trade agreements and tariff-based trade costs to explain economic integration in Africa. The trade cost (TC) in terms of tariffs is calculated as the average tariff a country imposes on imports from other African countries. This is obtained by summing the bilateral tariffs with each African trade partner and dividing by the number of partners (Vidya and Taghizadeh-Hesary, 2021). Higher trade costs are expected to reduce trade integration. We also use regional trade agreements (RTAs) as trade policies influencing African countries' connectivity or network position. According to Mon and Kakinaka's (2020) study, the impact of RTAs on each country depends on the country's economic size and the size of the markets accessible through its RTA partners, rather than solely on the number of accessible markets. The measure of regional trade agreements (RTAs) thus incorporates the economic sizes of each country and its RTA partners. Specifically, we compute the weighted measures of RTAs for country i represented by $RTA_{it} = \frac{1}{GDP_{it}} \sum_{j=1}^{J} B_{ij}\, GDP_{jt}$, where $B_{ij}$ is RTA tie dummies in which it assigns one if i and j in the same trade bloc and 0 otherwise, and $GDP_{it}$ and $GDP_{jt}$ represent the real GDP of country i and its RTA partner j.

Since many African countries belong to multiple RTAs, we adjust for potential double-counting using the overlap frequency ratio following Chacha (2014), which measures the proportion of a country's RTA memberships that overlap with other agreements. This ensures that the variable reflects the intensity of overlapping membership without overstating its effect. The ratio captures the frequency of overlapping memberships relative to total membership. This ratio focuses on the frequency of overlapping memberships rather than the sheer quantity of such memberships. Most African countries belong to more than one regional trading community. The Economic Commission for Africa suggests that overlapping membership is one of the major problems for African regional integration (AU, 2012; Zhang & Batinge, 2021). In our estimation, a global monetary crisis is treated as a dummy variable that takes the value 1 when a country was in a crisis (2007-2009) and 0 otherwise.

As a comparison exercise, we also report results using conventional de facto indicators of trade and financial integration. The most used measure of trade integration is the ratio of total imports and exports to GDP, which serves as a common proxy for openness to international markets. Financial integration is represented by a stock-based measure, the ratio of total foreign assets and liabilities to GDP, with data on financial integration obtained from (Lane et al., 2017). Following previous studies of Osei et al, (2019) and Ibrahim (2020) and Kim et al. (2022), we adopt de facto measures, which reflect actual trade and financial flows, rather than de jure measures of policy restrictions, for three main reasons: (i) they are less prone to endogeneity and better capture countries' real openness; (ii) they incorporate the actual effects of liberalization policies; and (iii) they provide greater cross-country and time variation in our dataset. These measures capture countries' foreign debt and net financial positions in explaining the integration process. In this specification, trade and financial integration[3] are proxied by conventional indicators, namely trade flows and asset inflows, respectively, rather than by network centrality measures.

**Estimation strategies and Endogeneity issues**

This study employs the System Generalized Method of Moments (System GMM) within a dynamic panel framework because countries' positions in the trade network are persistent over time. Including the lagged dependent variables creates correlation with unobserved country-specific effects, making pooled OLS and standard fixed-effects estimators biased in short dynamic panels. System GMM addresses this problem by combining equations in first differences and levels and by using lagged values of endogenous and predetermined variables as internal instruments (Arellano and Bover, 1995; Blundell and Bond, 1998). This technique improves efficiency relative to Difference GMM, particularly when the variables are persistent, as lagged levels may otherwise be weak instruments for differenced equations. By exploiting additional moment conditions, System GMM provides a more reliable identification strategy in short-panel settings with a large cross-sectional dimension. The lagged dependent variable is treated as predetermined, GDP per capita as endogenous, and the remaining controls according to their timing assumptions. To limit instrument proliferation, the instrument matrix is collapsed, and the lag depth is restricted following Roodman (2009). Country and year effects are included in all specifications.

Model validity is assessed using standard diagnostic tests. We report the Arellano–Bond test for second-order serial correlation in the differenced residuals to verify the absence of serial correlation in the error term. Instrument validity is evaluated using the Hansen test of overidentifying restrictions. In addition, to assess instrument relevance and guard against weak-instrument bias, we report Kleibergen–Paap LM and F statistics following the transformation approach proposed by Kripfganz (2019). Overall, these diagnostics support the appropriateness of the System GMM specification.

[3] When network centrality measures are used as explanatory variables in the trade and financial integration specifications, they may be endogenous because countries with stronger network positions may also attract greater financial inflows or trade volumes. We address this concern by using lagged instruments within the dynamic panel.

### 3.4. Data types and source

The analysis uses data from the United Nations Conference on Trade and Development between 2000 and 2019 (www.unctad.org). To construct the trade network, data for 52 African countries are used for the period 2000–2011. From 2012 onward, following the secession of South Sudan, the sample expands to 54 countries, reflecting the inclusion of both Sudan and South Sudan. The trade matrix is constructed using total bilateral import and export flows (in USD) over the period 2000–2019. A list of countries and their ISO-Code3 can be found in Table A.6 in the appendix. The macroeconomic variables are collected from the World Bank World Development Indicators (www.databank.worldbank.org), except for the human capital index, which is taken from the Penn World Table database (www.ggdc.net/pwt). The data for institutional quality comes from the World Development Indicators (WDI) series published by the World Bank. The bilateral FDI data are drawn from the IMF's Coordinated Direct Investment Survey (CDIS), which provides bilateral investment positions for African reporting economies. For economies that do not report, particularly many offshore financial centers, we rely on mirror data from their counterpart countries. The African country sample used to construct the FDI network varies across 2009–2019 and is dictated solely by data availability. To ensure relevance, only investment stocks exceeding USD 1 million each year are included. Due to the unavailability of certain macroeconomic variables, the econometric analysis was conducted using a panel dataset comprising only 40 African countries for the period 2002–2019, as data for some countries were missing. Bilateral trade is included when the value of trade between countries exceeds the first quartile of the distribution of bilateral trade flows, to reduce noise from sporadic or negligible trade relationships. Network measures are computed using the NetworkX package in Python, and econometric analysis is conducted using the R package.

Table 1: Variables description and data sources.

| Variable | Description | Data Source |
|---|---|---|
| **Network Centrality** | weighted in-degree ($S^{in}$), out-degree ($S^{out}$), PageRank (PR), betweenness (B), random-walk betweenness (RWB), closeness (C), clustering coefficient (CC), and k-core | network model |
| **RGDPc** | Real GDP per capita (constant prices) | World Bank (WDI) |
| **HC** | Human capital index (based on average years of schooling and returns-to-education) | Penn World Table (PWT) 10.10 |
| **POP** | Population (total, log-transformed) | World Bank (WDI) |
| **TC** | Trade cost (average tariff, normalized by the sum of countries) | UNCTAD, WITS |
| **Infra** | PCA-based index from telephone lines, air freight, energy use, and electricity consumption | World Bank (WDI) |
| **IQI** | Institutional quality index from PCA of six governance indicators | Worldwide Governance Indicators (WGI) |

| RTA | Weighted RTA index using partner countries' GDP and shared RTA ties | WTO RTA Database, IMF |
|---|---|---|
| **FDI** | Foreign Direct Investment inflows (% of GDP) | World Bank (WDI) |
| **OFR** | Overlapping Frequency Ratio of shared RTA memberships | WTO RTA Database |
| **Financial Crisis** | Dummy variable indicating financial crisis years (2007-2009) | Laeven & Valencia, IMF |

## 4. Results and Discussion

### 4.1. The African Trade Network

#### 4.1.1. Network Evolution

The evolution of the African trade network can be analyzed using a complex network framework that captures the structure and intensity of trade relationships among countries. This allows a system-wide assessment of economic integration beyond bilateral interactions. To evaluate changes in network structure, we use the network density, which reflects overall connectivity, and the average clustering coefficient, which captures the degree of local clustering among countries.

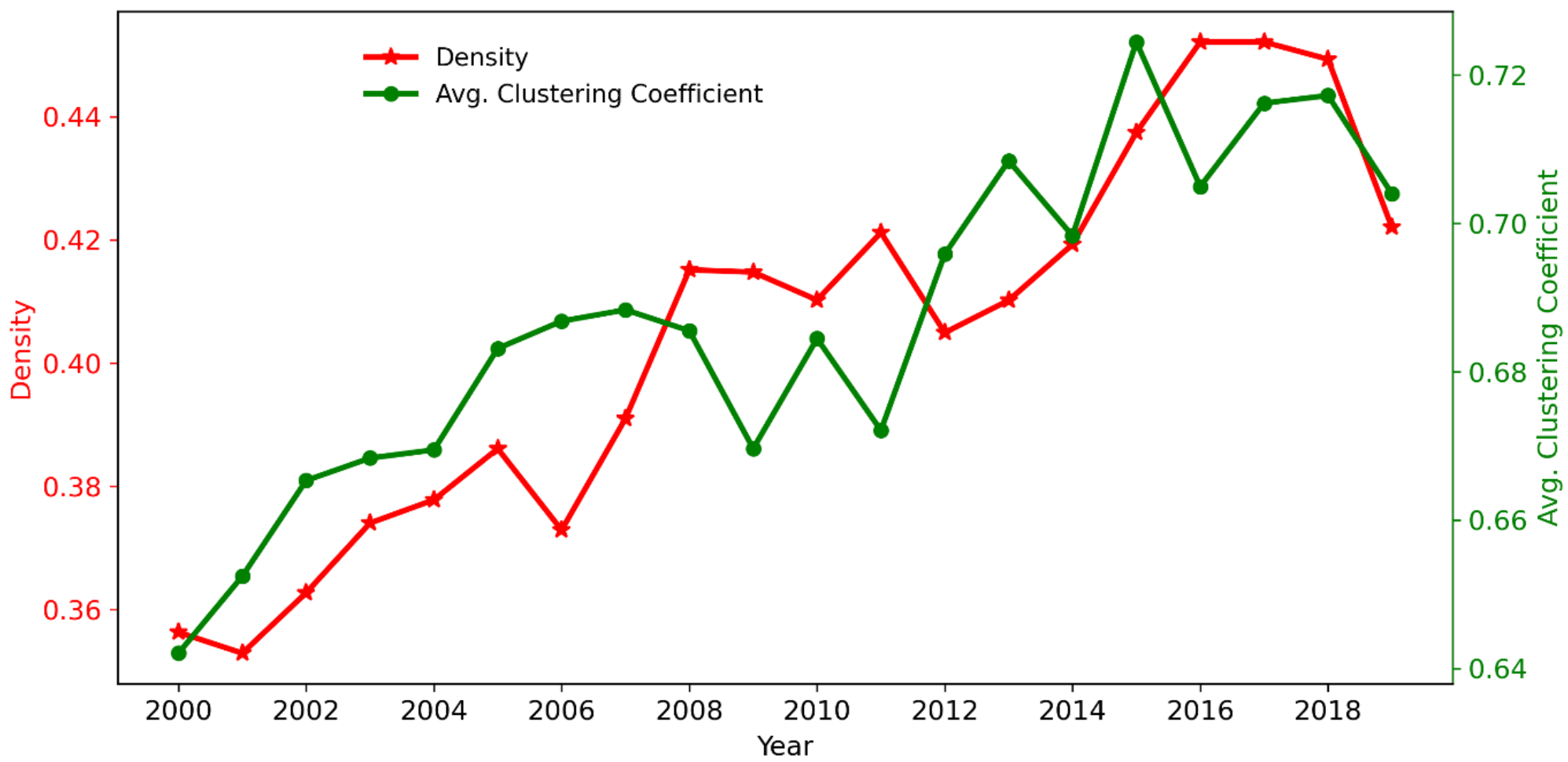

Figure 4: Evolution of structural characteristics of the African trade network. The panel shows a twin plot of density (left) and average clustering coefficient (right).

The upward trend in network density is consistent with a gradual strengthening of intra-African trade interactions during the 21st century (Figure 4). The density of intra-African trade ranges between 0.35 and 0.42, remaining below the density typically reported for the world trade network, which is around or above 0.5 in comparable studies. The average clustering coefficient also rises over time, indicating that African trade partners are increasingly connected to one another (Fig. 4). Overall, this means that intra-African trade remains only moderately interconnected, leaving potential for stronger economic integration.

#### 4.1.2. Evolution of Network Centrality Measures

Countries occupying central network positions play a central role in regional connectivity. Figure 5a shows that South Africa (ZAF), Egypt, Côte d'Ivoire, and Kenya (KEN) emerge as highly connected countries within the African trade network. These countries rank among the most integrated in terms of the number of trade partners (i.e., network degree). Figure 5b shows that South Africa, Kenya, Egypt (EGY), and Cote d'Ivoire (CIV) consistently held prominent positions in terms of PageRank.

Figure 5c shows that South Africa, Egypt, and Tunisia were the top-performing African countries in terms of bridging roles during the study period, with South Africa leading due to rapid infrastructural development and economic growth (See Table A.2 in the Supplementary Information). Changes in betweenness rankings over time indicate evolving brokerage roles within the African trade network. Nigeria, Tunisia, and Ghana exhibit lower prominence as bridging countries, suggesting a weaker brokerage role within the network. At the same time, Ethiopia and Uganda rose from 31st and 23rd in 2000 to 5th and 12th in 2019, respectively. The increase in trade connections appears to have reduced asymmetries in the network, weakening the intermediary role of some countries (Fig. 5c), consistent with the findings of Iapadre & Tajoli (2017).

The analysis of closeness centrality also shows that South Africa remained the most central country, with Egypt and Kenya having faster access to other countries, more resources, and greater influence (See Table A.2 in the Supplementary Information). Cote d'Ivoire, Tunisia, and Tanzania had the highest closeness centrality in some years, indicating that these countries have short trade paths and that their trade policies may quickly affect other countries. This suggests that these countries may have an advantage in obtaining resources or entering other countries' markets. Comoros, Eritrea, Lesotho, Cabo Verde, São Tomé and Príncipe, and Somalia shifted towards the periphery, while Seychelles moved away from a low position due to strong trade ties with influential countries like South Africa, Mauritius, Tunisia, Egypt, and Cote d'Ivoire.

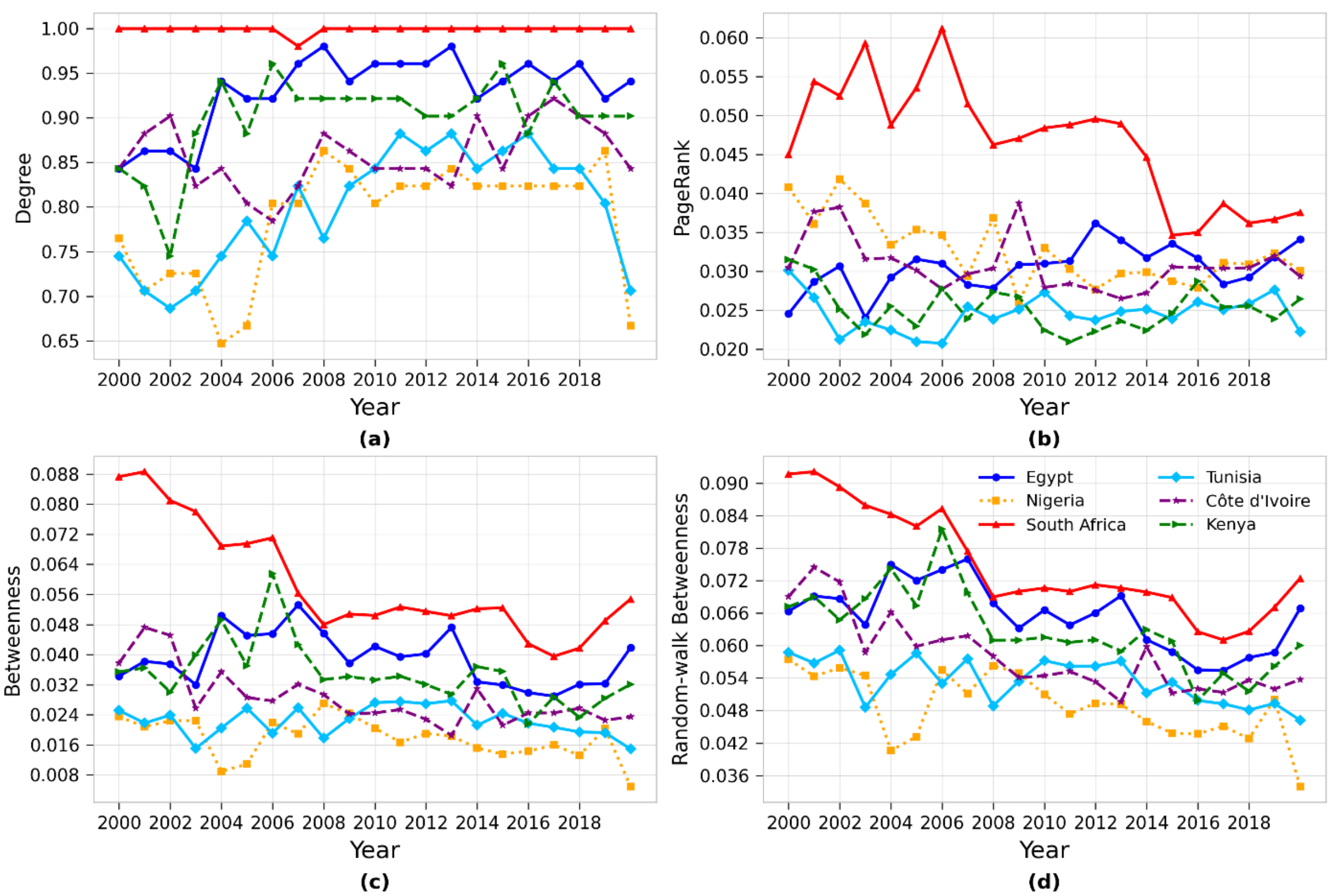

Figure 5: Network centrality measures of six selected African countries[4] from 2000-2019: (a) Degree; (b) PageRank; (c) Betweenness; (d) Random-walk betweenness centralities. The selected countries represent major economies across African subregions and account for a large share of continental GDP and population.

Figure 5 generally shows that initially dominant countries gradually declined in relative prominence. This pattern reflects a redistribution of trade linkages and the emergence of new regional hubs as intra-African integration deepened, accompanied by rising centrality scores among previously peripheral economies, signaling greater convergence and interconnectedness across the continent. Iapadre and Lucia Tajoli (2014) highlights that increasing global integration and broader geographic diversification of trade can reduce the relative concentration of intra-regional ties and push nodes toward a more dispersed network, lowering relative centrality in the regional subnetwork.

### 4.1.3. Evolution of Network Clustering Measures

Figure 6a shows that clustering in the trade network has increased over the past 20 years, suggesting stronger trade among countries with common partners. South Africa, Kenya, and

[4] Table A.2 in the appendix provides a list of the top and bottom ten countries in the trade network based on PageRank, betweenness, random walk, and closeness centrality.

Egypt had relatively low clustering coefficients throughout the study period, with increasing connectivity in recent years. Countries previously less embedded in tightly connected clusters increasingly formed links with partners inside those clusters. A country with only a few trading partners is more likely to trade frequently with those same partners because it struggles to enter new markets. This increases trade within that small group of countries, resulting in a higher clustering coefficient. On the other hand, some African countries with many trading partners tend to trade extensively within their existing group of partners and with other countries outside that group. The upward trend in clustering is consistent with increasing triadic closure and local interconnectedness in intra-African trade (Garlaschelli & Loffredo, 2005; Fagiolo et al., 2010; Cingolani et al., 2018).

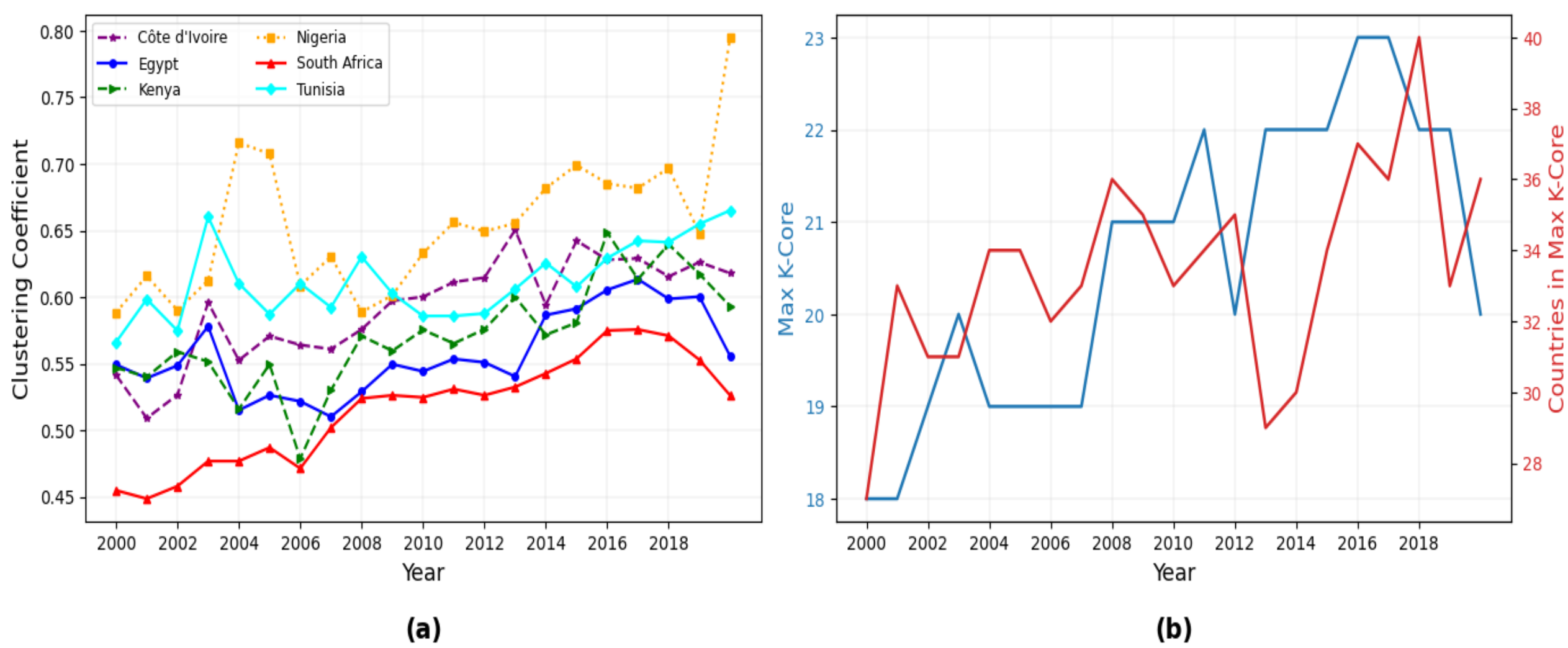

Figure 6: Network clustering measures of African countries from 2000-2019: (a) evolution of clustering coefficient for six selected African countries and (b) evolution of k-core. The selected countries represent major economies across African subregions and account for a large share of continental GDP and population.

Figure 6b shows an increasing trend in k-core, indicating a gradual strengthening of the network core over time. Higher k-core values indicate that countries are embedded in more densely interconnected trade substructures, consistent with deeper structural integration. The top-ranked countries in 2000 remain at the top in 2019, with slight changes in the k-core position (Fig. A.1 in the Supplementary Information). The African trade network exhibits a core-periphery structure. Countries with the largest k-core are highly interconnected and responsible for maintaining network connectivity, while those with the lowest k-core are peripheral. Over time, a larger share of countries appears in higher k-cores, consistent with an expansion of the network core (Fig. 6b). In 2018, Rwanda, Eswatini, and Ethiopia joined the maximum k-core, while Angola, Mozambique, Uganda, and Zambia lost their positions.

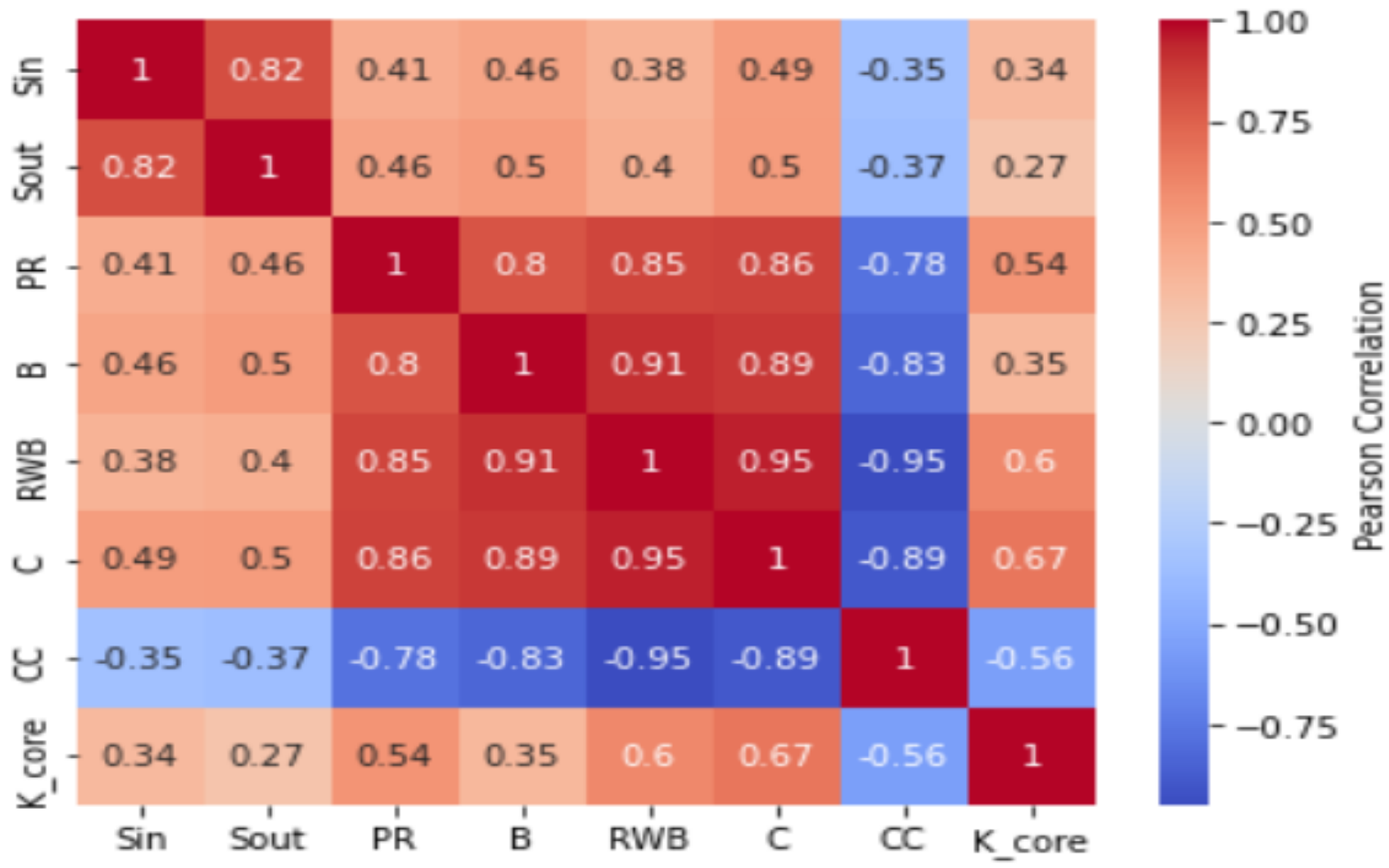

Figure 7: The Pearson correlation coefficient of network centrality measures taken for all countries (n=54) from 2000 to 2019.

Figure 7 shows that high correlations are observed between certain centrality indicators in dense and hierarchical trade networks. Because all measures are derived from the same adjacency matrix, countries occupying core positions tend to score highly across multiple centrality dimensions (Freeman, 1979; Borgatti, 2005; Newman, 2010). In core–periphery systems and networks shaped by preferential attachment, connectivity, influence, and embeddedness co-evolve, producing statistical covariance among indicators consistent with network theory by Borgatti & Everett (2000). However, each metric captures a distinct structural mechanism, local connectivity (degree), brokerage power (betweenness), recursive prestige (PageRank), or cohesive embeddedness (k-core and clustering), reflecting structural interdependence rather than dimensional redundancy (Figure 5 and Table A.2 in the appendix).

These cross-country differences illustrate that, even when indicators are statistically correlated, they represent distinct mechanisms of network positioning, direct connectivity (degree), intermediation power (betweenness), and recursive influence (PageRank). Therefore, correlation reflects partial structural overlap within a dense trade system but does not imply that the measures capture the same dimension of economic integration. Empirical studies of international trade networks similarly report substantial covariance among centrality metrics, while emphasizing that each indicator represents a different dimension of network organization (Fagiolo et al., 2010; De Benedictis & Tajoli, 2011). These patterns underscore the multifaceted nature of trade integration: some countries are broadly connected, others act as strategic intermediaries, and some derive influence through their embeddedness within highly central clusters even when empirical correlations are present.

Figure 7 shows the negative association between centrality indicators and clustering coefficients, which supports the presence of a core–periphery structure, where a core group of highly connected countries occupies central positions in the network, while peripheral countries remain more sparsely linked (Reyes et al., 2008; De Benedictis and Tajoli, 2011). The positive correlations between exporter and importer trade intensity measures (i.e., out- and in-degree) and among centrality metrics further underscore the importance of trade competitiveness. Countries with many inward edges typically also have many outward edges, suggesting strong bi-directional trade relationships and reinforcing the interconnected nature of the continent's trade dynamics.

To assess structural integration, we use MENM-normalized Z-scores of centrality measures. Specifically, we calculate PageRank, betweenness, and k-core annually on trade-weighted directed networks and normalize them using the maximum-entropy null model described in Section 3.2.1. The heat maps of PageRank, Betweenness, and k-core Z-scores from 2000 to 2019 illustrate the evolution of the relative importance of African economies over time (see Figure A.5, A.6 and A.7 in the appendix). A higher Z-score indicates that a country's observed centrality exceeds that of the null model, signaling a disproportionately influential or connected economy. A clear core-periphery structure emerges along all three indicators. A few economies dominate continental trade, while many others remain structurally peripheral. The observation that many African economies exhibit rising Z-scores over time is consistent with increased structural integration and supportive of Hypothesis H1.

South Africa, Egypt, Morocco, Nigeria, Côte d'Ivoire, and Kenya exhibit persistently high Z-scores, indicating sustained hub positions in intra-African trade. They combine high PageRank (influence) with high betweenness (brokerage), linking otherwise weakly connected parts of the network. Mid-tier economies like Ghana, Tanzania, and Senegal show gradually increasing Z-scores, indicating rising structural prominence over time. On the other hand, economies such as Burundi, Chad, and Guinea-Bissau maintain persistently negative Z-scores, indicating limited integration into the network core. Moreover, Figures A.2 and A.3 (in the appendix), showing the PageRank centrality against expected values, reveal persistent structural asymmetries within the African trade network. It highlights that certain economies consistently outperform their expected centrality levels, while others remain peripheral or underconnected throughout the two-decade period examined.

After 2015, k-core Z-scores rise for a broader set of economies, suggesting a widening of the network core. Such structural convergence suggests that even peripheral economies were beginning to form more stable intra-African linkages. The highest centrality concentration in a few key economies illustrates the broker-oriented nature of Africa's trade network, which remains core-centered.

### 4.2. Econometrics Analysis

Before estimating the dynamic specification, we report pooled OLS and fixed-effects estimates as benchmarks. We use the Hausman specification test to assess whether fixed effects are preferred to random effects. Detailed information on all diagnostic tests can be found in Table A.3 (Supplementary information). Benchmark pooled OLS and fixed-effects results are reported in Appendix Table A.8. Across specifications, GDP per capita is positively associated with most centrality measures, while higher trade costs are generally associated with lower trade intensity and weaker structural embeddedness. Several coefficients vary across centrality definitions and estimation methods, motivating the dynamic-system GMM specification reported below.

Table 2 reports the system GMM estimates of the impact of macroeconomic indicators on network centrality measures. To evaluate the reliability of the instruments and the dynamic panel model specification, several diagnostic tests were conducted. The system GMM estimates in Table 2 support the conclusion that the instruments are both valid and relevant. In all columns, where the dependent variables capture different dimensions of network centrality (columns 1–8), the Kleibergen–Paap F-statistic is above the Stock and Yogo (2005) critical values, for the null that the IV estimates' bias exceeds 10% of the OLS bias, thereby confirming the strength and relevance of the instruments for these specifications. Furthermore, the Kleibergen–Paap LM statistic rejects the null of underidentification, confirming that the models are adequately identified. The Hansen J-statistic does not reject the null hypothesis that the overidentifying restrictions are valid, providing strong evidence for the instruments' joint exogeneity. Arellano–Bond tests indicate first-order serial correlation in differenced residuals, as expected, and no evidence of second-order serial correlation. To avoid sample bias from overfitting, the number of instruments was limited to at most the number of groups in the regression. In all specifications, the number of instruments remains below the number of country groups.

Table 2. System GMM estimation for network-based integration measures

| **Variable** | $S^{in}$ | $S^{out}$ | **PR** | **B** | **RWB** | **C** | **cc** | **k-core** |
|---|---|---|---|---|---|---|---|---|
| **RGDPc** | 0.255*** | 0.0702*** | 0.0150 ** | 0.0785** | 0.0074* | 0.0031** | -0.0015*** | 0.635*** |
| | (0.016) | (0.0206) | (0.005) | (0.039) | (0.0042) | (0.0014) | (0.001) | (0.056) |
| **HC** | 0.107** | 0.0816 | 0.0199** | -0.0180 | 0.0071 | -0.0042** | 0.0048** | 0.751*** |
| | (0.053) | (0.0762) | (0.010) | (0.0796) | (0.0111) | (0.0018) | (0.0024) | (0.217) |
| **POP** | 0.032** | 0.0602*** | -0.006** | 0.0487* | 0.0063*** | 0.0014 | 0.0004 | -1.0864** |
| | (0.013) | (0.0209) | (0.003) | (0.0273) | (0.0025) | (0.0010) | (0.0008) | (0.1291) |
| **TC** | -0.021*** | -0.0436*** | -0.0041 | -0.094** | -0.0084 | -0.0017*** | 0.0004 | -1.362*** |
| | (0.0044) | (0.0119) | (0.0125) | (0.0454) | (0.0085) | (0.0006) | (0.0005) | (0.255) |
| **Infra** | 0.002 | 0.0003 | 0.0021** | 0.0327** | -0.0013 | -0.0004 | 0.0084*** | 1.351*** |
| | (0.008) | (0.007) | (0.0008) | (0.0121) | (0.0009) | (0.0003) | (0.0043) | (0.519) |
| **IQI** | -0.0012 | 0.0286** | 0.0044 | 0.0073 | 0.0026 | -0.0003 | 0.0004** | 1.007*** |
| | (0.012) | (0.0150) | (0.0032) | (0.0149) | (0.0019) | (0.0006) | (0.0002) | (0.346) |
| **RTAs** | 0.0002* | -0.00014 | 0.0003*** | 0.0003 | - 0.0003*** | 0.0001 | -0.0003 | 1.821*** |
| | (0.0001) | (0.0001) | (0.0001) | (0.0002) | (0.0001) | (0.0002) | (0.0006) | (0.653) |
| **FDI** | 0.0175 | 0.0078*** | -0.0120 | 0.1729** | 0.0013 | 0.0314** | -0.0116*** | 1.435 |
| | (0.0892) | (0.0029) | (0.034) | (0.0372) | (0.0145) | (0.0132) | (0.0046) | (0.876) |
| **OFR** | -0.158*** | 0.0880 | -0.0194 | -0.2230* | 0.0251** | -0.0054 | -0.0082** | -1.043 |
| | (0.076) | (0.1135) | (0.0134) | (0.1340) | (0.0119) | (0.0034) | (0.0033) | (0.64) |
| **Financial crisis** | 0.242*** | -0.2168 | -0.0070 | -0.0452 | - 0.0221** | 0.0106*** | -0.004** | -1.332*** |
| | (0.021) | (0.0688) | (0.014) | (0.1048) | (0.0103) | (0.0035) | (0.002) | (0.572) |
| **L.$S^{in}$** | 0.889*** | | | | | | | |
| | (0.0179) | | | | | | | |
| **L.$S^{out}$** | | 0.911*** | | | | | | |
| | | (0.0164) | | | | | | |
| **L.PR** | | | 0.961 ** | | | | | |
| | | | (0.014) | | | | | |
| **L.B** | | | | 0.888***<br>(0.038) | | | | |
| **L.RWB** | | | | | 0.747***<br>(0.151) | | | |
| **L.C** | | | | | | 0.890***<br>(0.249) | | |
| **L.Cc** | | | | | | | 0.966***<br>(0.0219) | |
| **L.k-core** | | | | | | | | 0.922**<br>(0.188) |
| Hansen test | 33.317 | 4.470 | 20.717 | 4.105 | 27.717 | 10.77 | 8.211 | 27.717 |
| Hansen p-value | 0.0976 | 0.613 | 0.899 | 0.6625 | 0.542 | 0.291 | 0.6082 | 0.458 |
| AR(1) test | -14.329 | 3.670 | 25.213 | 4.092 | 12.272 | 2.484 | -4.974 | 12.272 |
| AR(1)p-value | 0.0001 | 0.0002 | 0.0000 | 0.000 | 0.0000 | 0.013 | 0.0000 | 0.0000 |
| AR(2) test | 0.1445 | 1.003 | 0.782 | 0.171 | 0.741 | 0.937 | 1.412 | 0.741 |
| AR(2)p-value | 0.889 | 0.316 | 0.184 | 0.864 | 0.118 | 0.349 | 0.1159 | 0.904 |
| **Instruments** | | | | | | | | |
| Number of IVs | 35 | 33 | 32 | 18 | 33 | 22 | 35 | 40 |
| IVs for the 1st Diff | L(2/5).X | L(2/4).X | L(2/4).X | L(2/4).X | L(2/5).X | L(2/4).X | L(2/4).X | L(2/7).X |
| IVs for the level | D.X | D.X | D.X | D.X | D.X | D.X | D.X | D.X |
| Underidentification test | | | | | | | | |
| Kleibergen-Paap LM stats | 45.724 | 36.091 | 40.808 | 20.240 | 35.049 | 31.934 | 39.331 | 40.491 |
| p-value | 0.0317 | 0.0401 | 0.037 | 0.009 | 0.050 | 0.0014 | 0.0491 | 0.000 |
| Weak identification test | | | | | | | | |
| Kleibergen-Paap Wald stats | 443.78 | 157.418 | 90.79 | 57.496 | 137.353 | 16.815 | 243.897 | 26.374 |
| Stock-Yogo critical values | | | | | | | | |
| 10% maximal IV relative bias | 11.06 | 11.05 | 11.05 | 10.43 | 11.05 | 10.84 | 11.05 | 9.37 |
| 20% maximal IV relative bias | 6.03 | 6.06 | 6.08 | 6.22 | 6.06 | 6.21 | 6.01 | 5.78 |
| 30% maximal IV relative bias | 4.29 | 4.32 | 4.35 | 4.69 | 4.32 | 4.56 | 4.26 | 4.46 |
| Wald test | 0.001 | 0.0000 | 0.0063 | 0.0022 | 0.0179 | 0.000 | 0.0002 | 0.0002 |
| Obs | 680 | 680 | 680 | 680 | 680 | 680 | 680 | 680 |

*Notes: Clustered robust standard errors in parentheses. *** p<0.01, **p<0.05 and * p<0.10.* All specifications include country and year effects. The lagged dependent variable is treated as predetermined, while GDP per capita is treated as endogenous. The instrument matrix is collapsed to reduce instrument proliferation following Roodman (2009). The Windmeijer finite-sample correction is applied to the two-step System GMM standard errors.

Table 2 also presents the estimation results from the dynamic panel model using the generalized method of moments (GMM). Network centrality measures are highly persistent over time.

Accordingly, we include the lagged dependent variable in the dynamic specification. The lagged dependent variable is positive and statistically significant across specifications, indicating strong persistence in countries' network positions. This persistence implies that improvements in network position tend to carry over into subsequent years.

The coefficient on RGDP per capita is positive and statistically significant across all network-based integration measures except the clustering coefficient, for which the relationship is negative. Specifically, a 1% increase in RGDP per capita is associated with a 0.015% increase in PageRank centrality. This supports hypothesis H2. Economically, higher GDP per capita is associated with stronger, more influential positions within the intra-African trade network. This finding is consistent with existing evidence linking economic size to more central positions in trade networks (Pédussel Wu, 2004; Garlaschelli and Loffredo, 2005; Fagiolo et al., 2009; Beaton et al., 2017; Akbari et al., 2021; Herman, 2022). Conversely, the negative coefficient of GDP per capita on the clustering coefficient does not contradict agglomeration theory, but rather reflects a structural transition in the organization of trade networks. As countries develop, they tend to move from regionally dense trade clusters toward more diversified, hub-like continental integration. In network terms, higher-income countries occupy core or bridging positions, connecting to heterogeneous partners that are not mutually connected, thereby reducing local clustering. This pattern is consistent with hierarchical and core–periphery network structures observed in international trade.

The results indicate that human capital is positively associated with PageRank, clustering, and k-core, while its association with brokerage-related measures is weaker. This suggests that improvements in education, skills, and knowledge help countries connect to influential partners and become embedded in cohesive trade communities. Economies with stronger human capital are likely to have greater productive capacity, better absorptive capabilities, and stronger institutional and managerial knowledge, which can support repeated trade relationships within regional value chains. However, human capital alone does not necessarily turn a country into a broker between otherwise weakly connected parts of the network. Brokerage positions depend not only on domestic capabilities but also on geography, transport corridors, historical trade routes, and the wider layout of trade pathways. Thus, human capital appears to deepen integration within the network core without necessarily shortening all trade paths or creating new intermediation roles across the continent.

Table 2 reports that population size appears to be an important determinant of a country's position within the trade network, particularly for indicators capturing intermediary (bridging) roles and the intensity of direct trade connections. However, the results indicate a negative association between population size and PageRank centrality, an indicator of economic influence derived from trade linkages with highly connected and economically important partners, as well as with k-core, which captures structural embeddedness within densely interconnected trade clusters. This finding suggests that demographic size alone does not automatically translate into greater economic influence or deeper core integration within the intra-African trade network. This pattern is consistent with highly populated countries being less structurally embedded in the continental core, despite having large domestic markets. As a result, they play a less central and less structurally embedded role within the African trade network. At the same time, population size is positively associated with weighted in-degree and out-degree, suggesting that countries with larger populations maintain a high intensity of trade connections with their partners, even if their structural importance in the network is limited.

Our findings show a positive link between institutional quality and both weighted out-degree and cluster-based trade centrality measures at conventional levels of significance. By contrast, institutional quality is not statistically significant for brokerage-related measures. This pattern can be explained by the distinct structural requirements of different network positions. Strong institutions—such as effective contract enforcement, regulatory quality, and governance stability—reduce transaction costs and uncertainty in bilateral trade relationships. This facilitates the formation and maintenance of direct trade ties, thereby increasing a country's trade volume with multiple partners (weighted degree) and reinforcing participation within tightly interconnected regional clusters (clustering). The positive relationship with clustering measures suggests that countries with stronger institutions are better able to integrate into dense groups of trading partners because institutional strength fosters trust, transparency, and lower trade costs. Institutional quality is also positively associated with k-core, consistent with greater embeddedness in the network core. This interpretation aligns with prior work linking institutional quality to lower trade frictions and stronger trade ties (Jiang & Tamang, 2020; Zhang et al., 2021). Conversely, betweenness centrality captures a country's role as an intermediary or bridge connecting otherwise disconnected parts of the network. Such brokerage positions are determined not by institutional quality alone, but rather by geographic location, transport corridors, historical trade routes, regional trade agreements, and the overall structure of the trade network. Accordingly, improvements in institutional quality may strengthen direct trade connections without necessarily altering a country's strategic intermediary position. The empirical finding that institutional quality significantly affects direct trade relationships but not betweenness centrality is therefore consistent with the theoretical distinction between direct trade connectivity and network brokerage roles. Therefore, while strong institutions promote deeper and broader direct trade integration, they do not necessarily guarantee that a country becomes a key intermediary between other trading partners.

As hypothesized, regional trade agreements (RTAs) are among the most important determinants of a country's position in the trade network. RTAs have a significant positive effect on influence-related measures, including PageRank and weighted in-degree, reflecting their role in enhancing a country's integration, visibility, and overall influence within the network. This empirical evidence is consistent with Hypothesis H2. In contrast, RTAs show a negative association with random-walk centrality, suggesting that participation in regional trade agreements does not necessarily make countries common transit hubs in African trade; instead, it may concentrate trade within specific blocs, reducing their role in wider continental trade flows. As a result, countries experience higher visibility and influence (reflected in PageRank centrality) and attract more direct trade volume (captured by weighted Indegree), thereby improving their overall centrality within the network. RTAs are also positively associated with k-core, indicating stronger embeddedness in the network core. The findings are consistent with empirical literature (Jiménez-García and Rodríguez, 2022; Sada et al., 2022; Kang et al., 2023). However, their impact is less pronounced for betweenness, closeness, and clustering, suggesting that while RTAs enhance connectivity and embeddedness, they do not necessarily increase a country's control over trade flows or local network density.

Infrastructure quality is positively associated with several measures of network-based economic integration. Better infrastructure enhances countries' positions in terms of influence (PageRank), brokerage (Betweenness centrality), and core embeddedness (clustering coefficient and k-core) (Table 2). This association is consistent with the idea that infrastructure reduces frictions in cross-border trade. This can lead to higher trade flows and greater connectivity, which can contribute to the country's higher network position (Francois and Manchin, 2013; Shepherd, 2016; Vidya and Taghizadeh-Hesary, 2021). We do not find robust effects on random-walk and closeness centrality once controls are included.

Trade costs, measured as the simple average bilateral tariff rates among African countries, exhibit a negative and economically significant association with countries' centrality in the African trade network. Higher trade costs reduce countries' influence and trade competitiveness (as reflected in weighted out-degree and weighted in-degree), weaken their intermediary and bridging roles (as captured by betweenness centrality), and limit their structural embeddedness within cohesive trade clusters (as indicated by k-core). Thus, higher tariffs are associated with weaker trade intensity, reduced brokerage roles, and lower embeddedness in the network core. This finding aligns with previous studies showing that trade costs, such as tariffs and non-tariff measures, can be considered barriers to trade and reduce countries' network positions (Novy, 2013; Basile et al., 2018). The underlying theory of Krugman (1991) states that increased spatial concentration of production and trade integration follow reductions in trade costs. However, we find that for closeness and the clustering coefficient, the estimated association with tariffs is positive. This result may reflect that tariffs are correlated with other unobserved trade-policy or network-structure characteristics in these measures; we therefore interpret these coefficients with caution.

The coefficient of FDI shows a positive impact on all network centrality indicators except import trade volume of partners, random-walk centrality, and k-core, suggesting that inward FDI generally enhances a country's strategic position in the trade network. This effect can be explained by the role of multinational corporations in expanding production capacity, fostering technology transfer, and improving infrastructure provision, which collectively strengthen trade connectivity. However, FDI is negatively associated with the clustering coefficient, indicating that while foreign investment deepens bilateral ties and broadens linkages, it may reduce the density of localized trade clusters by reorienting flows toward global rather than regional partners. Positive spillovers, therefore, help strengthen a country's central positions within the network. Moreover, the significant positive interaction terms between FDI and RTA and exporter trade intensity, betweenness, PageRank, closeness, and k-core indicate that the presence of regional trade agreements is critical for attracting foreign investment and enhancing trade integration (see Table A.9 in the Supplementary Information). On the other hand, more regional trade agreements (RTAs) among African countries have boosted foreign direct investment (FDI), leading to more complex economic interactions and improved economic integration.

The overlapping frequency ratio, which captures the extent of overlapping trade memberships, is found to have a detrimental effect on network centrality. Specifically, higher overlapping membership significantly reduces countries' weighted indegree, weakens intermediary positions (betweenness), lowers their likelihood of being frequently reached within trade pathways (random-walk centrality), and diminishes their participation in cohesive trade clusters (clustering coefficient), all at conventional significance levels. A country with overlapping regional memberships loses its position in the network; i.e., African countries with more than one REC hinder trade integration, reducing the country's in-degree because multiple memberships entail different rules and regulations for meeting trade cooperation requirements. The difficulty of establishing regulatory structures within such an economic bloc may explain the low level of intra-African trade. Regional membership can promote integration when rules are coherent, but overlapping memberships may fragment incentives and raise compliance costs. AfCFTA may help reduce these frictions if it simplifies and harmonizes rules across regional blocs (Zhang & Batinge, 2021; Haiyun et al., 2023).

Table 3 reports that under the difference-GMM specifications, the AR(2) and Hansen diagnostics comfortably exceed conventional thresholds, supporting the validity of the moment conditions. However, the instrument relevance is inadequate for each integration measure in Table 4, the Kleibergen–Paap F-statistic falls short of the Stock–Yogo (2005) critical values, indicating weak identification. We thus interpret the diff-GMM coefficients with caution and rely on the system-GMM estimates for inference.

Table 3. GMM estimates for determinants of the de facto measure of economic integration in the context of financial and trade integration.

| Variable | difference GMM | | System GMM | |
|---|---|---|---|---|
| | TI | FI | TI | FI |
| **RGDPc** | 0.243<br>(0.205) | 0.0023***<br>(0.005) | 0.0414**<br>(0.0221) | 0.0563**<br>(0.0241) |
| **HC** | 0.115**<br>(0.066) | 0.0016<br>(0.060) | 0.0054<br>(0.0043) | -0.0230<br>(0.0344) |
| **POP** | 0.022<br>(0.067) | 0.0322<br>(0.0217) | -0.0260<br>(0.0138) | 0.0567*<br>(0.0332) |
| **TC** | -0.056**<br>(0.0225) | 0.0045<br>(0.0039) | -0.0076***<br>(0.0025) | -0.0553<br>(0.0436) |
| **Infra** | 0.0072<br>(0.0052) | 0.0004<br>(0.0054) | 0.0055<br>(0.0047) | 0.0006<br>(0.0029) |
| **IQI** | -0.0045<br>(0.034) | 0.0205**<br>(0.0101) | 0.0039<br>(0.0030) | 0.0079**<br>(0.0049) |
| **RTAs** | 0.0143**<br>(0.0073) | -0.0439**<br>(0.0221) | 0.0153**<br>(0.0061) | 0.0065**<br>(0.0033) |
| **OFR** | -0.147<br>(0.099) | 0.087<br>(0.063) | -0.068**<br>(0.0034) | -0.133**<br>(0.034) |

| Financial crisis | -0.325<br>(0.226) | -0.117<br>(0.088) | -0.272<br>(0.426) | -0.102<br>(0.095) |
|---|---|---|---|---|
| **Hansen test** | 25.105 | 20.717 | 27.07 | 23.53 |
| **Hansen p-value** | 0.275 | 0.509 | 0.306 | 0.253 |
| **AR(2) test** | 0.049 | 0.782 | 0.199 | 0.792 |
| **AR(2) p-value** | 0.944 | 0.184 | 0.711 | 0.194 |
| **Instruments** | | | | |
| **Number of IVs** | 36 | 32 | 32 | 28 |
| **IVs for the 1st diff** | L(2/5).X | L(2/4).X | L(2/4).X | L(2/3).X |
| **IVs for the levels** | | | D.X | D.X |
| **Underidentification test** | | | | |
| **Kleibergen-Paap LM stat** | 45.724 | 36.091 | 38.143 | 38.130 |
| **p-value** | 0.0303 | 0.0340 | 0.040 | 0.005 |
| **Weak identification test** | | | | |
| **Kleibergen-Paap Wald stats** | 3.366 | 4.608 | 5.263 | 3.465 |
| **Stock-Yogo critical values** | | | | |
| **10% maximal IV relative bias** | 10.72 | 10.75 | 10.77 | 10.76 |
| **20% maximal IV relative bias** | 5.92 | 5.97 | 5.93 | 5.98 |
| **30% maximal IV relative bias** | 4.24 | 4.39 | 4.24 | 4.39 |
| **Observation** | 680 | 680 | 680 | 680 |

Notes: Robust standard errors are reported in parentheses. *** $p < 0.01$, ** $p < 0.05$, and * $p < 0.10$.

Table 4 shows that several macroeconomic variables are not statistically significant for the traditional trade and financial integration measures. Overlapping membership and GDP are significant for both trade and financial integration, institutional quality is significant for financial integration, and trade costs are significant for trade integration. This suggests that conventional indicators may capture a narrower set of correlates of economic integration in Africa. When comparing network-based measures of financial integration, such as betweenness, PageRank, clustering coefficient, and k-core, to traditional measures like the trade-to-GDP ratio or the ratio of total foreign assets and liabilities to GDP, we find they provide complementary information on countries' structural positions. Empirically, the network-based specifications yield more stable coefficient patterns across alternative integration dimensions and highlight mechanisms—such as brokerage and core embeddedness—that are not visible in aggregate ratios. The comparison suggests that network-based metrics offer a more nuanced view of integration patterns than aggregate ratio measures, particularly in the African context.

Table 4: System GMM estimates of the impact of network-based measures on the *De facto* measure of trade and financial integration.

| **Variable** | **Dependent variable: Trade integration** | | | | **Dependent variable: Financial integration** | | | |
|---|---|---|---|---|---|---|---|---|
| | **(1)** | **(2)** | **(3)** | **(4)** | **(5)** | **(6)** | **(7)** | **(8)** |
| **B** | 0.056**<br>(0.023) | | | | 0.0188<br>(0.0156) | | | |
| **PR** | | 0.0305**<br>(0.016) | | | | 0.0145**<br>(0.0078) | | |
| **C** | | | -0.0034<br>(0.0045) | | | | -0.0170**<br>(0.0098) | |
| **K-core** | | | | 0.5265 **<br>(0.2102) | | | | 0.658***<br>(0.255) |
| **RGDPc** | 0.0419**<br>(0.0220) | 0.0433**<br>(0.0206) | 0.0465**<br>(0.0204) | 0.0436**<br>(0.0217) | 0.0178*<br>(0.0047) | 0.0462**<br>(0.0221) | 0.0306**<br>(0.0141) | 0.0215**<br>(0.0101) |
| **HC** | 0.0059<br>(0.0044) | 0.0056<br>(0.0048) | 0.0057<br>(0.0045) | 0.0055<br>(0.0048) | -0.0260<br>(0.0355) | -0.0270<br>(0.0343) | -0.0253<br>(0.0332) | -0.0275<br>(0.0305) |
| **POP** | -0.0253<br>(0.0138) | -0.0232<br>(0.0138) | -0.0271<br>(0.0138) | -0.0260<br>(0.0135) | 0.0543<br>(0.0344) | 0.0525<br>(0.0345) | 0.0557<br>(0.0366) | 0.0518<br>(0.0339) |
| **TC** | -0.0066**<br>(0.0031) | -0.0065**<br>(0.0029) | -0.0063***<br>(0.0023) | -0.0064**<br>(0.0028) | -0.0465<br>(0.0402) | -0.0556<br>(0.0435) | -0.0515<br>(0.0493) | -0.0519<br>(0.0416) |
| **Infra** | 0.0046<br>(0.0071) | 0.0050<br>(0.0069) | 0.0049<br>(0.0055) | 0.0048<br>(0.0057) | 0.0005<br>(0.0039) | 0.0006<br>(0.0035) | 0.0005<br>(0.0030) | 0.0007<br>(0.0048) |
| **IQI** | 0.0033<br>(0.0035) | 0.0036<br>(0.0028) | 0.0035<br>(0.0025) | 0.0034<br>(0.0032) | 0.0080**<br>(0.0041) | 0.0073**<br>(0.0036) | 0.0075**<br>(0.0037) | 0.0077**<br>(0.0038) |
| **RTAs** | 0.0144**<br>(0.0060) | 0.0123**<br>(0.0059) | 0.0125**<br>(0.0063) | 0.0145**<br>(0.0062) | 0.0065**<br>(0.0033) | 0.0065**<br>(0.0033) | 0.0065**<br>(0.0033) | 0.0065**<br>(0.0033) |
| **OFR** | -0.061**<br>(0.0028) | -0.062**<br>(0.0030) | -0.064**<br>(0.0033) | -0.058**<br>(0.0025) | -0.130**<br>(0.039) | -0.128**<br>(0.042) | -0.131**<br>(0.041) | -0.127**<br>(0.044) |
| **Financial crisis** | -0.232<br>(0.323) | -0.206<br>(0.320) | -0.253<br>(0.316) | -0.232<br>(0.293) | -0.115<br>(0.090) | -0.112<br>(0.089) | -0.109<br>(0.093) | -0.107<br>(0.087) |
| **Hansen test** | 28.07 | 27.67 | 20.07 | 18.07 | 23.88 | 19.45 | 25.53 | 20.53 |
| **Hansen p-value** | 0.307 | 0.606 | 0.417 | 0.523 | 0.258 | 0.303 | 0.357 | 0.359 |
| **AR(2) test** | 0.358 | 0.887 | 0.992 | 0.345 | 0.345 | 0.476 | 0.169 | 0.307 |
| **AR(2)p-value** | 0.715 | 0.206 | 0.190 | 0.644 | 0.722 | 0.634 | 0.869 | 0.725 |
| **Instruments** | | | | | | | | |
| **Number of IVs** | 32 | 32 | 32 | 32 | 35 | 32 | 32 | 36 |
| **IVs for the 1st Diff** | L(2/4).X | L(2/4).X | L(2/4).X | L(2/4).X | L(2/5).X | L(2/4).X | L(2/4).X | L(2/7).X |
| **IVs for the levels** | D.X | D.X | D.X | D.X | D.X | D.X | D.X | D.X |
| **Kleibergen-Paap LM stat** | 40.247 | 39.218 | 36.364 | 42.095 | 38.055 | 32.404 | 40.394 | 37.461 |

| p-value | 0.050 | 0.033 | 0.0025 | 0.026 | 0.043 | 0.0018 | 0.0458 | 0.000 |
|---|---|---|---|---|---|---|---|---|
| Kleibergen-Paap Wald stat | 55.254 | 80.69 | 36.638 | 70.860 | 86.341 | 66.657 | 57.828 | 37.336 |
| 10% maximal IV relative bias | 10.46 | 11.08 | 10.90 | 11.06 | 11.06 | 10.81 | 11.05 | 10.38 |
| 20% maximal IV relative bias | 6.23 | 6.08 | 6.24 | 6.09 | 6.07 | 6.24 | 6.02 | 6.08 |
| 30% maximal IV relative bias | 4.65 | 4.36 | 4.55 | 4.36 | 4.33 | 4.55 | 4.26 | 4.37 |
| Obs | 680 | 680 | 680 | 680 | 680 | 680 | 680 | 680 |

Notes: Robust standard errors are reported in parentheses. *** p < 0.01, ** p < 0.05, and * p < 0.10. X denotes the set of instrumented variables, including lagged trade integration, lagged financial integration, GDP per capita, and the relevant network centrality indicator.

We regress the selected measures of financial integration, betweenness, PageRank, clustering coefficient, and k-core on the de facto measures of trade and financial integration. The analysis focuses on trade openness (TI) and asset-liability flow (FI), which are standard de facto indicators with consistent coverage in the sample. To address potential endogeneity, the lagged values of the independent variables are employed as instruments. Table 5 indicates that network-based measures are significantly associated with de facto trade and financial integration, and the coefficient signs align with the main results in Table 2. Several macroeconomic covariates retain their expected signs under the traditional integration measures. The positive relationships observed are consistent with earlier findings in Tables 2 and 3, with GDP, human capital, and overlapping membership retaining their significance and expected signs.

The findings suggest that trade openness and cross-border asset–liability flows play pivotal roles in Africa’s trade and financial integration. Moreover, network connectedness is systematically related to these de facto indicators, underscoring the value of incorporating network measures alongside traditional indicators. As a robustness check, we implement an alternative IV strategy following Niu et al. (2018) (See Table A.10 in the appendix). While lagged regressors help address reverse causality concerns, we also consider an external-instrument approach as a robustness exercise. To address the issue, we treated the intensity of a country's exports to its top 5 trading partners as an external instrument. This instrument is intended to proxy external trade exposure while remaining plausibly predetermined with respect to contemporaneous shocks. Prior evidence suggests export intensity is correlated with key macroeconomic fundamentals (Jalles, 2012). We examine the interaction between trade intensity and the remaining endogenous macroeconomic variables and use the interaction term as an instrument for each factor. The main coefficient signs are broadly consistent with the baseline estimates, and key results remain statistically supported in the robustness specification.

## 5. Conclusion and policy implications

The main objective of this study is to introduce and empirically examine network-based measures of economic integration and to assess how macroeconomic fundamentals shape countries' position in the African trade network. Unlike traditional measures of integration, we proposed advanced network measures that capture higher-order interdependencies in the African trade network. We link these network-based integration measures to macroeconomic indicators using a dynamic panel regression framework.

The analysis of the African trade network shows that intra-African trade has expanded and become denser, reflecting greater diversification of trade partners across the continent. Several countries, including Ethiopia, Angola, and Uganda, have moved from peripheral positions toward more central roles in the trade network. On the other hand, countries such as South Africa, Egypt, Côte d'Ivoire, Kenya, Morocco, Senegal, Tanzania, and Ghana consistently occupy central positions across multiple network indicators, reflecting strong and diversified intra-African trade linkages.

We found that the African trade network contains a core-periphery structure, and the trade agglomeration effect of the core countries has increased over time. Countries with high k-core values, located at the network's core, are more exposed to system-wide shocks and may transmit disturbances more widely (Askari et al, 2018). We also confirmed our hypothesis that common partners of African countries are increasingly trading with each other (i.e., increasing the clustering coefficient). As a result, highly central countries play influential roles in the diffusion of goods and resources across the continent and may amplify the transmission of economic shocks.

The econometric analysis identifies key macroeconomic factors associated with countries' structural positions in the African trade network. Higher economic development, participation in regional trade agreements, institutional quality, human capital, and infrastructure are positively associated with more central network positions. The results suggest that countries improve their structural position in the trade network through two related channels: stronger domestic fundamentals, such as income, human capital, institutions, and infrastructure, and lower external frictions, such as trade costs and overlapping regional commitments. Improving infrastructure quality is essential for enhancing a country's centrality and increasing its access to resources and markets. Therefore, investment in infrastructure development could be a crucial policy intervention to enhance a country's competitiveness in African trade. Lowering tariffs and non-tariff trade barriers is also crucial for reducing trade costs. Overlapping regional memberships are negatively associated with several network dimensions, particularly k-core, clustering, and random-walk centrality, indicating weaker structural embeddedness.

The findings offer several implications for understanding economic integration in Africa and for interpreting the conditions under which AfCFTA may strengthen continental trade. Because the sample ends in 2019, the analysis should be read as a pre-implementation structural baseline rather than as an evaluation of AfCFTA's causal effects. Countries participating in regional integration should coordinate economic policy adjustments, reduce transaction costs, and simplify overlapping regional commitments to improve the structure of continental connectivity. Policymakers should further prioritize enhancing economic performance, institutional quality, trade openness, and human capital, as these factors are crucial in strengthening trade networks, as indicated by k-core, clustering coefficient, PageRank, and betweenness centrality. Leveraging these network-based measures can not only optimize the structure and expansion of trade across Africa but also provide valuable insights to mitigate the effects of economic and financial shocks, ensuring more resilient and interconnected economies. Finally, overlapping membership in various regional economic groups can hinder economic integration, as many African countries face non-tariff barriers, such as sourcing rules and quality requirements that impede trade. Multiple membership is associated with increased transaction costs, a main challenge in Africa (African Union, 2012).

This study measures economic integration from a complex systems perspective, capturing countries' structural roles, i.e., hubness, brokerage, and embeddedness, within the African trade system. By using network-based measures instead of aggregate ratios alone, the study shows that integration is also a question of structural position, i.e., who connects to whom, who bridges otherwise weakly connected parts of the continent, and who belongs to the trade core. This study focuses on the intra-African trade network. The overlap between this network and the global trade network remains outside the scope of the analysis and may change countries' relative positions. Future research should examine sector-specific trade networks and regional economic blocs separately, as these would help distinguish trade complementarity from substitutability among African economies. From this perspective, AfCFTA should be evaluated not only by whether it increases intra-African trade shares, but also by whether it changes the structure of African trade by connecting peripheral economies, widening the continental core, and reducing excessive dependence on a few dominant hubs.

## Acknowledgments

The authors thank Glenn Rayp for his comments on the manuscript. T.T. and L.E.C.R. thank the financial support of the Network for Advancement of Sustainable Capacity in Education and Research in Ethiopia (NASCERE programme), coordinated by the University of Ghent, Belgium, and Jimma University, Ethiopia. T.T thanks financial support from the Bijzonder Onderzoeksfonds (BOF) (2025/0108/01) from Ghent University, Belgium L.E.C.R. is partially funded by the Bijzonder Onderzoeksfonds (BOF) (2024/01/709) from Ghent University, Belgium.